\documentstyle[aps,prl,multicol,fancyheadings,rotate,epsfig,amsbsy,amssymb,amstex]{revtex}
\def\one{{\mathchoice {\rm 1\mskip-4mu l} {\rm 1\mskip-4mu l} {\rm
1\mskip-4.5mu l} {\rm 1\mskip-5mu l}}}
\def\bbbc{{\mathchoice {\setbox0=\hbox{$\displaystyle\rm C$}\hbox{\hbox
to0pt{\kern0.4\wd0\vrule height0.9\ht0\hss}\box0}}
{\setbox0=\hbox{$\textstyle\rm C$}\hbox{\hbox
to0pt{\kern0.4\wd0\vrule height0.9\ht0\hss}\box0}}
{\setbox0=\hbox{$\scriptstyle\rm C$}\hbox{\hbox
to0pt{\kern0.4\wd0\vrule height0.9\ht0\hss}\box0}}
{\setbox0=\hbox{$\scriptscriptstyle\rm C$}\hbox{\hbox
to0pt{\kern0.4\wd0\vrule height0.9\ht0\hss}\box0}}}}

\begin{document}
\title{Generalized Jordan-Wigner Transformations}
\author{C.D. Batista and G. Ortiz}
\address{Theoretical Division, 
Los Alamos National Laboratory, Los Alamos, NM 87545}
\date{Received \today }
\maketitle

\begin{abstract}
We introduce a new spin-fermion mapping, for arbitrary spin $S$
generating the $SU(2)$ group algebra, that constitutes a natural
generalization of the Jordan-Wigner transformation for $S=\frac{1}{2}$.
The mapping, valid for regular lattices in any spatial dimension $d$,
serves to unravel hidden symmetries in one representation that are
manifest in the other. We illustrate the power of the transformation by
finding exact solutions to lattice models previously unsolved by
standard techniques. We also present a proof of the existence of the
Haldane gap in $S=$1 bilinear nearest-neighbors Heisenberg spin chains
and discuss the relevance of the mapping to models of strongly
correlated electrons. Moreover, we present a general spin-anyon mapping
for the case $d \leq 2$.
\end{abstract}

\pacs{Pacs Numbers: 75.10.Jm, 03.65.Fd, 05.50.+q, 11.10.-z}

\vspace*{-0.4cm} 

\begin{multicols}{2}

\columnseprule 0pt

\narrowtext
\vspace*{-0.5cm}	

{\it Introduction.}
Theories of magnetism in normal matter are direct manifestations of
quantum mechanics. The vast majority of phenomena occurring in magnets,
such as different types of magnetically ordered states, are often
simply described by interacting quantum spins \cite{mattis}. Although
quantum spins do not behave as either pure boson or fermion operators,
different  representations have invoked such particle statistics, for
example, the Holstein-Primakoff, Schwinger, and Dyson-Maleev boson
representations for arbitrary spin $S$, or the Jordan-Wigner and
Majorana fermion representations for spin $S$=$\frac{1}{2}$ magnets. 

Sometimes it is extremely convenient to reformulate a difficult
strongly correlated problem in a way that it becomes more manageable;
in some cases there is an exact dualism. This is the idea behind the 
bosonization techniques and the different algebra representations of a
physical problem. The point is that these different representations
help us understand various aspects of the same problem by transforming
intricate interaction terms into simpler ones. Often, fundamental
symmetries which are hidden in one representation are manifest   
in the other and, moreover, problems which seem untractable can even be
exactly solved after the mapping. The simplest and perhaps most popular
example is the equivalence between the Heisenberg-Ising
$S$=$\frac{1}{2}$ XXZ chain and a model of interacting spinless
fermions through the Jordan-Wigner transformation \cite{jordan}.
 
The Jordan-Wigner transformation involves the $S$=$\frac{1}{2}$
irreducible representation of the Lie group $SU(2)$. Here we generalize
this spin-fermion mapping to any irreducible representation of
dimension $2S+1$. The three generators $S_j^\mu (\mu=x,y,z)$ of the Lie
algebra for each lattice site $j$ satisfy the commutation relations
\cite{Note0}
\begin{equation}
\left [ S_j^\mu, S_k^\nu \right ] = i \delta_{j k} \epsilon_{\mu \nu
\lambda} S_j^\lambda \ ,
\end{equation}
with $\epsilon$ the totally antisymmetric tensor. The algebra
generated by the (linear and Hermitian operators) $S_j^\mu$ 
is the enveloping algebra of the group $SU(2)$. Equivalently, in terms
of the ladder operators $S_j^\pm = S_j^x \pm i S_j^y$
\begin{eqnarray}
\left [ S_j^+, S_j^- \right ] &=& 2 S_j^z \ ,   
\left [ S_j^z, S_j^\pm \right ] = \pm S_j^\pm \ , \nonumber \\
\left \{ S_j^+, S_j^- \right \} &=& 2 \left( S(S+1)- (S_j^z)^2 \right )
\ .
\end{eqnarray}
We start by analyzing the one-dimensional $S$=1 case. Then, we will
show a generalization to arbitrary spin and spatial dimension $d$.

{\it $S$=1 Mapping.} 
We introduce the following composite operators
\begin{equation}
f_j^\dagger = \bar{c}^\dagger_{j 1} + \bar{c}^{\;}_{j \bar{1}} \ , \ 
f_j^{\;} = \bar{c}^{\;}_{j 1} + \bar{c}^\dagger_{j \bar{1}} \ , 
\end{equation}
written in terms of the Hubbard operators $\bar{c}^\dagger_{j \sigma}=
c^\dagger_{j \sigma} (1 - \bar{n}_{j \bar{\sigma}})$ and
$\bar{c}^{\;}_{j \sigma}= (1 - \bar{n}_{j \bar{\sigma}}) \, c^{\;}_{j
\sigma}$ ($\sigma=1,-1$), which form a subalgebra of the so-called
double graded algebra $Spl(1,2)$ \cite{tsvelick}. [A bar in a subindex
means the negative of that number (e.g., $\bar{\sigma}=-\sigma$).] For
spins on a lattice we fermionize the spins and reproduce the correct
spin algebra with the following transformation
\begin{eqnarray}
S^+_j &=& \sqrt{2} \ (\bar{c}^\dagger_{j 1} \ K_j + K_j^\dagger \
\bar{c}^{\;}_{j \bar{1}}) \ , \nonumber \\ 
S^-_j &=& \sqrt{2} \ (K_j^\dagger \ \bar{c}^{\;}_{j 1} +
\bar{c}^\dagger_{j \bar{1}} \ K_j) \ , \nonumber \\ 
S^z_j &=& \bar{n}_{j 1} - \bar{n}_{j \bar{1}} \ ,
\nonumber 
\end{eqnarray}
whose inverse manifests the nonlocal character of the mapping
\begin{eqnarray}
f^\dagger_j &=& \frac{1}{\sqrt{2}} \ \exp[i \pi \sum_{k<j} (S_k^z)^2] \ 
S_j^+  \ , \nonumber \\ 
f^{\;}_j &=& \frac{1}{\sqrt{2}} \exp[-i \pi \sum_{k<j} (S_k^z)^2] \ S_j^-
\ , \nonumber \\
\bar{c}^\dagger_{j 1}&=& S_j^z \ f^\dagger_j \;\;\;\, , \ \ \
\bar{c}^{\;}_{j 1} =  f^{\;}_j \ S_j^z \ , \nonumber \\
\bar{c}^\dagger_{j \bar{1}}&=& - S_j^z \ f^{\;}_j \ , \ \;\;\, \bar{c}^{\;}_{j
\bar{1}} = -f^\dagger_j \ S_j^z \ , \nonumber 
\end{eqnarray}
where the string operators $K_j = \exp[i \pi \sum_{k<j}\bar{n}_k] =
\prod_{k<j} \prod_\sigma (1 - 2 \bar{n}_{j \sigma})$, and the number
operators $\bar{n}_k = \bar{n}_{k 1} + \bar{n}_{k \bar{1}}$. These
$f$-operators have the remarkable property that
\begin{equation}
\left \{ f_j^\dagger, f_j^{\;} \right \} =  \left \{ S_j^+, S_j^- \right
\} \ ,
\end{equation}
which suggests an analogy between spin operators and ``constrained''
fermions. 

Half-odd integer spin chains have a qualitatively different excitation
spectrum than integer spin chains. The Lieb, Schultz, Mattis and
Affleck theorem \cite{spinp} establishes that the half-odd integer
antiferromagnetic (AF) bilinear nearest-neighbors (NN) Heisenberg chain
is gapless. The same model with integer spins is conjectured to
display a Haldane gap \cite{haldane}. To understand the origin of the
Haldane gap we analyze the form of the 1$d$ $S$=1 XXZ Hamiltonian using
the above representation (an overall omitted constant $J$ $>$ 0
determines the energy scale) 
\begin{eqnarray}
H_{\rm xxz} &=& \sum_j S^z_j S^z_{j+1} + \Delta \left ( S^x_j S^x_{j+1}
+ S^y_j S^y_{j+1} \right ) \nonumber \\
&=& \sum_j H_j^z + H_j^{\rm xx} \ .
\end{eqnarray}
It is easy to show that the constrained fermion version of this
Hamiltonian is a ($S=$$\frac{1}{2}$) $t$-$J_z$ model \cite{nos1} plus a
particle non-conserving term which breaks the $U(1)$ symmetry
\begin{eqnarray}
H_{\rm xxz} &=& \sum_j (\bar{n}_{j 1} - \bar{n}_{j
\bar{1}})(\bar{n}_{j+1 1} - \bar{n}_{j+1 \bar{1}}) \nonumber \\
&+& \Delta \sum_{j
\sigma} \left ( \bar{c}^\dagger_{j \sigma} \bar{c}^{\;}_{j+1 \sigma} +
\bar{c}^\dagger_{j \sigma} \bar{c}^\dagger_{j+1 \bar{\sigma}} + {\rm
H.c.} \right ) \ .
\end{eqnarray}
In the isotropic $\Delta=1$ limit, $H_{\rm xxz}$ can be written in a
compact way
\begin{eqnarray}
H_{\rm Heisenberg} &=& \sum_j \left( \Psi^\dagger_j \vec{\bf S} \Psi^{\;}_j
\right) \cdot \left(  \Psi^\dagger_{j+1} \vec{\bf S} \Psi^{\;}_{j+1}
\right) \ ,
\label{heisenberg}
\end{eqnarray}
where $\vec{\bf S}$ is an irreducible matrix representation of $S$=1
(3$\times$3 matrices) while $\Psi^\dagger_j$ is the ($1 \times 3$)
vector 
\begin{equation}
\Psi^\dagger_j = \begin{bmatrix}
         \bar{n}_{j 1} \!\!\!&,& \!\! 
	 (\bar{c}^\dagger_{j 1}+ \bar{c}^\dagger_{j \bar{1}}) K_j \!\!\!&,&\!\!
         \bar{n}_{j \bar{1}} 
         \end{bmatrix} \ .
\end{equation}
 
The charge spectrum of the ($S=\frac{1}{2}$) $t$-$J_z$ model is gapless
but the spin spectrum is gapped due to the explicitly broken $SU(2)$
symmetry (Luther-Emery liquid) \cite{nos1}. Therefore, the spectrum of
the $S=$1 Hamiltonian associated to the $t$-$J_z$ model (which has only
spin excitations) is gapless. Hence the term which explicitly breaks
$U(1)$ must be responsible for the opening of the Haldane gap. We can
prove this by  considering the perturbative effect that the interaction
$\eta \sum_{j \sigma} (\bar{c}^\dagger_{j \sigma}\bar{c}^\dagger_{j+1
\bar{\sigma}} + {\rm H.c.})$ has on the $t$-$J_z$ Hamiltonian. To
linear order in $\eta$ ($>0$), Eq. \ref{heisenberg} maps onto the
($S=\frac{1}{2}$) XYZ model with ${\cal J}_x=2+\eta$, ${\cal
J}_y=2-\eta$, and ${\cal J}_z=-1$. From exact solution of this model
\cite{Baxter}, it is seen that the system is critical only when
$\eta=0$ while for $\eta \neq 0$ a gap to all excitations opens. 

{\it $S$=1 Integrable Models}\cite{cristian}.
To illustrate further the power of our spin-fermion mapping we now
present {\it exact} solutions of 1$d$ $S$=1 models that have not been
discovered by traditional techniques. These models correspond to the
family of bilinear-biquadratic Hamiltonians 
\begin{eqnarray}
H_1(\Delta) &=& \sum_j H_j^z + H_j^{\rm xx} + \left \{ H_j^z, H_j^{\rm xx}
\right \} \nonumber \\
&=& \sum_j H_j^z + \Delta \sum_\sigma \bar{c}^\dagger_{j \sigma}
\bar{c}^{\;}_{j+1 \sigma} \ ,
\end{eqnarray}
that can be mapped onto a ($S=$$\frac{1}{2}$) $t$-$J_z$ Hamiltonian, whose
quantum phase diagram has recently been exactly solved \cite{nos1}.

Another well-studied class of bilinear-biquadratic $SU(2)$ invariant
Hamiltonians is \cite{solyom}
\begin{equation}
H_2(\Delta) = \sum_j {\bf S}_j \cdot {\bf S}_{j+1} + \Delta \left (
{\bf S}_j \cdot {\bf S}_{j+1} \right )^2 \ ,
\label{spin}
\end{equation}
for -1$\leq$ $\Delta$ $\leq$1. The pure Heisenberg ($\Delta$=0) and
Valence Bond Solid models ($\Delta=\frac{1}{3}$) belong to the Haldane
gapped phase, which extends over the whole interval except at the
boundaries $\Delta=\pm 1$ that are quantum critical points. The case
$\Delta$=-1 is known to be Bethe ansatz soluble with a unique ground
state and gapless. For $\Delta$=1 we can map $H_2(1)$ onto the
supersymmetric ($S=$$\frac{1}{2}$) $t$-$J$ Hamiltonian plus a NN
repulsive interaction 
\begin{eqnarray}
H_2(1)&=&-\sum_{j \sigma} \left ( \bar{c}^\dagger_{j \sigma}
\bar{c}^{\;}_{j+1 \sigma} + {\rm H.c.} \right ) + 2 \sum_j {\bf
s}_j \cdot {\bf s}_{j+1} \nonumber \\
&+& 2 \sum_j (1-\bar{n}_j + \frac{3}{4} \bar{n}_j \bar{n}_{j+1}) \ ,
\label{carga}
\end{eqnarray}
where ${\bf s}_j$ represents a $S=$$\frac{1}{2}$ operator. This model
is Bethe-ansatz soluble with a gapless phase \cite{pedro} and is known
as the Lai-Sutherland solution \cite{lai}.

We now discuss the importance of our generalized Jordan-Wigner
transformation in unraveling hidden symmetries of an arbitrary spin
Hamiltonian. In Eq. \ref{spin}, for example, the $S$=1 $SU(2)$ symmetry
is manifest. However, both the $S$=$\frac{1}{2}$ $SU(2)$ and global
$U(1)$ gauge symmetries are hidden. On the other hand, in the
transformed Hamiltonian, Eq. \ref{carga}, these two symmetries are
manifested explicitly through rotational invariance and charge
conservation. The generators of these symmetries are related through
the mapping already introduced. To illustrate this, we consider the
$U(1)$ symmetry case. Here the generator of the transformation is
$Q=\sum_j \bar{n}_{j}$ which maps onto $Q=\sum_j (S_j^z)^2$ in the spin
representation. The total group symmetry of the Hamiltonian is $U(3)$. 

{\it Generalized Transformation.}
A general transformation for arbitrary spin and spatial dimension is
the following

\noindent
\underline{\bf Half-odd integer spin $S$} ($\sigma \in {\cal
F}_{\frac{1}{2}} = \{-S+1, \dots, S \}$): 

\begin{eqnarray}
S^+_j &=& \eta_{\bar{S}} \ \bar{c}^\dagger_{j \bar{S}+1} \ K_j + \sum
\Sb \sigma \in {\cal F}_{\frac{1}{2}} \\ \sigma \neq S \endSb \!
\eta_\sigma \ \bar{c}^\dagger_{j \sigma+1} \bar{c}^{\;}_{j \sigma}
\ , \nonumber \\ 
S^-_j &=& \eta_{\bar{S}} \ K_j^\dagger \ \bar{c}^{\;}_{j \bar{S}+1} +
\sum \Sb \sigma \in {\cal F}_{\frac{1}{2}} \\ \sigma \neq S \endSb \!
\eta_\sigma \ \bar{c}^\dagger_{j \sigma} \ \bar{c}^{\;}_{j \sigma+1}
\ , \nonumber \\ 
S^z_j &=& -S + \sum \Sb \sigma \in {\cal F}_{\frac{1}{2}} \endSb
(S+\sigma) \ \bar{n}_{j \sigma} \ , \nonumber \\
\bar{c}^\dagger_{j \sigma} &=& K_j^\dagger L_\sigma^{\frac{1}{2}} \left( S^+_j
\right)^{\sigma+S} {\cal P}_j^{\frac{1}{2}} \ , \nonumber \\
\mbox{where} && {\cal P}_j^{\frac{1}{2}} =
\!\! \prod \Sb \tau \in {\cal F}_{\frac{1}{2}} \endSb
\! \frac{\tau -S^z_j}{\tau + S} \ , \ L_\sigma^{\frac{1}{2}} =
\!\!\!\!\! \prod_{\tau=-S}^{\sigma-1} \!\!\! \eta_\tau^{-1} \ . \nonumber
\end{eqnarray}

\begin{figure}[htb]
\epsfverbosetrue
\epsfxsize=7cm
\centerline{\epsfbox{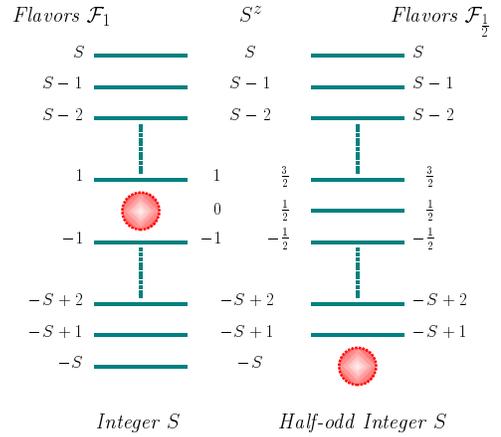}}
\vspace*{-0.8cm}
\caption{Constrained fermion states per site for integer and half-odd
integer spin $S$. In both cases there are $2S$ flavors and the 
corresponding $2S+1$ values of $S^z$ are shown in the middle column.
One degree of freedom is assigned to the fermion vacuum (circle) whose
relative position depends upon the spin character.}
\label{fig1}
\end{figure}

\noindent
\underline{\bf Integer spin $S$} ($\sigma \in {\cal F}_1 =
\{-S,\dots,-1,1,\dots,S \}$): 
 
\begin{eqnarray}
S^+_j &=& \eta_0 \ (\bar{c}^\dagger_{j 1} \ K_j + K_j^\dagger \
\bar{c}^{\;}_{j \bar{1}}) +  \sum \Sb \sigma \in {\cal F}_1 \\ \sigma \neq
-1,S \endSb \! \eta_\sigma \ \bar{c}^\dagger_{j \sigma+1}
\bar{c}^{\;}_{j \sigma} \ , \nonumber \\ 
S^-_j &=& \eta_0 \ (K_j^\dagger \ \bar{c}^{\;}_{j 1} +
\bar{c}^\dagger_{j \bar{1}} \ K_j) + \sum \Sb \sigma \in {\cal F}_1 \\ \sigma
\neq -1,S \endSb \! \eta_\sigma \ \bar{c}^\dagger_{j \sigma} \
\bar{c}^{\;}_{j \sigma+1} \ , \nonumber \\ 
S^z_j &=& \sum \Sb \sigma \in {\cal F}_1 \endSb \sigma \
\bar{n}_{j \sigma} \ , \nonumber \\
\bar{c}^\dagger_{j \sigma} &=& K_j^\dagger L_\sigma^1 \begin{cases}
            \left( S^+_j \right)^\sigma {\cal P}_j^1 & 
	    \text{if $\sigma > 0$} \ , \\
            \left( S^-_j \right)^\sigma {\cal P}_j^1 & 
	    \text{if $\sigma < 0$} \ , 
            \end{cases} \nonumber \\
\mbox{where} && {\cal P}_j^1 =
\!\! \prod \Sb \tau \in {\cal F}_1 \endSb
\! \frac{\tau -S^z_j}{\tau} \ , \ L_\sigma^1 =
\! \prod_{\tau=0}^{|\sigma|-1}  \eta_\tau^{-1} \ , \nonumber
\end{eqnarray}
and $\eta_\sigma = \sqrt{(S-\sigma)(S+\sigma+1)}$ (see Fig. \ref{fig1}). 

The total number of flavors is $N_f=2 S$, and the $S$=$\frac{1}{2}$ case 
simply reduces to the traditional Jordan-Wigner transformation. These
mappings enforce the condition on the Casimir operator ${\bf S}_j^2 =
S(S+1)$. The generalized constrained fields 
\begin{equation}
\bar{c}^\dagger_{j \sigma} = {c}^\dagger_{j \sigma} \!\!\! \prod \Sb
\tau \in {\cal F}_\alpha \endSb \!\! (1-{n}_{j
\tau}) \ , \ 
\bar{c}^{\;}_{j \sigma} = \!\!\!\! \prod \Sb \tau \in {\cal F}_\alpha
\endSb  \!\!(1-{n}_{j \tau}) \ {c}^{\;}_{j \sigma}
\ 
\end{equation}
form a subalgebra of the generalized Hubbard double graded algebra,
where the ``unconstrained'' operators $c^\dagger_{j \sigma}, c^{\;}_{j
\sigma}$ satisfy the standard fermion anticommutation relations 
($\alpha=\frac{1}{2},1$ depending upon the spin character of the
representation). These generalized constrained operators (only single
occupancy is allowed) anticommute for different sites
\begin{eqnarray}
\left \{ \bar{c}^{\;}_{j \sigma}, \bar{c}^{\;}_{k \sigma'} \right \} &=&
\left \{ \bar{c}^\dagger_{j \sigma}, \bar{c}^\dagger_{k \sigma'} \right
\} = 0  \ ,\nonumber \\
\left \{ \bar{c}^{\;}_{j \sigma}, \bar{c}^\dagger_{k \sigma'} \right \}
&=& \delta_{jk} \begin{cases}
\mbox{$\displaystyle \prod \Sb \tau \in {\cal F}_\alpha \\ \tau
                \neq \sigma \endSb  \!\!(1-\bar{n}_{j \tau})$} &
		\text{if $\sigma = \sigma'$}, \\
                \bar{c}^\dagger_{j \sigma'} \bar{c}^{\;}_{j \sigma} & 
		\text{if $\sigma \neq \sigma'$} , 
		\end{cases} 
\end{eqnarray}
and their number operators satisfy $\bar{n}_{j \sigma}\bar{n}_{j
\sigma'}= \delta_{\sigma \sigma'} \bar{n}_{j \sigma}$.

The string operators $K_j$ introduce nonlinear and nonlocal
interactions between the constrained fermions. For 1$d$ lattices ($K_j
= K_j^\dagger$, $\left[K_i, K_j\right]=0$) they are the so-called kink
operators $K_j= \exp[i \pi \sum\limits_{k < j} \bar{n}_k]$,
while for 2$d$ \cite{eduardo}
\begin{eqnarray}
K_{\bf j} &=& \exp[i \sum \Sb {\bf k} \endSb a({\bf k},{\bf j}) 
\ \bar{n}_{\bf k}] \ , \;\; \mbox{with} 
\label{string} \\
\bar{n}_{\bf k} &=& \sum \Sb \sigma \in {\cal F}_\alpha \endSb 
\bar{n}_{{\bf k} \sigma} = 1- {\cal P}_{\bf k}^\alpha \ .
\end{eqnarray}
Here, $a({\bf k},{\bf j})$ is the angle between the spatial vector
${\bf k}-{\bf j}$ and a fixed direction on the lattice, and $a({\bf
j},{\bf j})$ is defined to be zero. We comment that the 1$d$ kink
operators constitute a particular case of Eq. \ref{string} with
$a(k,j)=\pi$ when $k<j$ and equals zero otherwise. For $d>2$, the
string operators generalize \cite{cristian} along the lines
introduced in Ref. \cite{huerta}. 
\vspace*{-2.0cm}

There is always the freedom to perform rotations in spin space to get
equivalent representations to the one presented above. However, for
bilinear isotropic NN Heisenberg (spin $SU(2)$ rotationally invariant)
Hamiltonians in the large-$S$ limit there is a fundamental difference
between effective integer and half-odd integer spin cases. In the
latter case a new local $U(1)$ gauge symmetry emerges that is
explicitly broken in the integer case. For 1$d$ lattices, this is
precisely what distinguishes Haldane gap systems \cite{haldane} from
half-odd integer spin chains that are critical. 

We mention that other fermionic representations are feasible. In
particular, for half-odd integer cases where $2S+1 =
\sum_{i=0}^{\bar{N}_f} {\binom{\bar{N}_f}{i}}= 2^{\bar{N}_f}$ (e.g.,
$S$=$\frac{3}{2}$ with $\bar{N}_f=2$) a simple transformation in terms
of standard ``unconstrained'' fermions is possible \cite{cristian}. 

{\it 2$d$ Lattices and Spin-Anyon Mapping.}
The generalization of these transformations to higher dimensions gives
new exact mappings between spin theories and constrained fermion
systems in the presence of gauge fields. To illustrate this we write
the $S=$1 Hamiltonian $H_2(1)$ in the fermion representation for $d=2$ 
\begin{eqnarray}
H_2(1)&=&- \! \sum_{{\bf j} \sigma,\nu} \left ( \bar{c}^\dagger_{{\bf
j}+{\bf e}_\nu \sigma} \  e^{iA_{\nu}({\bf j})} \ \bar{c}^{\;}_{{\bf j}
\sigma} + {\rm H.c.} \right ) + 2 \sum_{{\bf j},\nu} {\bf s}_{\bf j}
\cdot {\bf s}_{{\bf j}+{\bf e}_\nu} \nonumber \\
&+& 2 \sum_{\bf j} (1-\bar{n}_{\bf j} + \frac{3}{4} \bar{n}_{\bf j}
\bar{n}_{{\bf j}+{\bf e}_\nu}) \ ,
\label{carga2}
\end{eqnarray}
and
\begin{equation}
A_{\nu}({\bf j})=\sum_{\bf k}[a({\bf k},{\bf j})-a({\bf k},{\bf j}+{\bf
e}_\nu)] \ \bar{n}_{\bf k} \ ,
\end{equation}
where ${\bf e}_\nu$ ($\nu=1,2$) are basis vectors of the Bravais
lattice connecting NN and ${\bf j}$'s represent sites of the
corresponding 2$d$ lattice. We note that the field $A_{\nu}({\bf j})$
is associated with the change in particle statistics. It is well-known
\cite{eduardo,tsvelick} that the same transmutation of particle
statistics can be achieved via a  path-integral formulation for
$H_2(1)$ where an Abelian lattice Chern-Simons term is included. In
this formulation a constraint (Gauss's law) requiring that the gauge
flux through a plaquette ${\bf j}$ be proportional to the total fermion
density on the site, $\bar{n}_{\bf j}$, is enforced.  This suggests
that our spin-fermion mapping can be generalized to an spin-anyon
transformation with a hard-core condition for the anyon fields
\cite{cristian}. In fact, one can formally take our generalized
Jordan-Wigner transformation and replace the string operators $K_{\bf
j}$ by the statistical operators $K_{\bf j}(\theta) = \exp[i \theta
\sum \Sb {\bf k} \endSb a({\bf k},{\bf j}) \bar{n}_{\bf k}]$ with $0
\leq \theta \leq 1$. With this choice, the $\bar{c}$ operators satisfy
equal-time anyon commutation relations [$\theta =1(0)$ corresponds to
constrained fermions(bosons)] \cite{cristian}. Similar ideas apply for
1$d$ lattices. 

One immediately sees the relevance of these transformations for the
theories of magnetism and high-temperature superconductivity: A class
of $S=$1 Hamiltonians that can be mapped onto a lattice-gauge
(Chern-Simons) $S=\frac{1}{2}$ $t$-$J$ theory and vice versa, for
example, a $S=\frac{1}{2}$ $t$-$J$ model,  
\begin{eqnarray}
H_{t-J} &=&-t\sum_{{\bf j} \sigma,\nu} \left ( \bar{c}^\dagger_{{\bf j}
\sigma} \bar{c}^{\;}_{{\bf j}+{\bf e}_\nu \sigma} + {\rm H.c.} \right )
+ J \sum_{{\bf j},\nu} {\bf s}_{\bf j} \cdot {\bf s}_{{\bf j}+{\bf
e}_\nu} \nonumber \\
&-& \mu \sum_{\bf j} \bar{n}_{\bf j}  \ ,
\label{carga3}
\end{eqnarray}
can be exactly mapped onto a lattice-gauge bilinear-biquadratic $S=$1
theory
\begin{eqnarray}
&H&_{t-J} = -\mu \sum_{{\bf j}} (S^z_{\bf j})^2 + \frac{J}{8}\sum_{{\bf
j},\nu} \left [ H^z_{{\bf j} \nu} \! -
\! \frac{4t}{J} S^+_{\bf j} e^{iA_{\nu}({\bf j})} S^-_{{\bf j}+{\bf
e}_\nu} \right . \nonumber \\
&-& \!\! \left . \frac{4t}{J} \left \{ H^z_{{\bf j} \nu}, S^+_{\bf j}
e^{iA_{\nu}({\bf j})} S^-_{{\bf j}+{\bf e}_\nu}\right \} + (S^+_{\bf j}
S^-_{{\bf j}+{\bf e}_\nu} )^2 + {\rm H.c.} \right ] \ .
\label{carga4}
\end{eqnarray}

By means of a semiclassical approximation it has been shown
\cite{Papanicolaou} that the ground state of $H_2(1)$ is on the 
boundary between AF ($\Delta < 1$) and orthogonal
nematic (non-uniform, $\Delta > 1$) phases \cite{Papanicolaou,solyom}.
These two states are the result of the competition between the
quadratic and quartic spin-exchange interactions. In terms of the
equivalent $t$-$J$ gauge theory this translates into a competition
between antiferromagnetism and delocalization. Qualitatively, the
string-path of the particle moving in an AF background
gives rise to a linear confining potential since the number of
frustrated magnetic links is proportional to the length of the path.
This observation suggests that the inhomogeneous phases observed in the
``striped'' high-T$_c$ compounds can be driven by the competition
between magnetism and delocalization. 
 
{\it Summary.}
We introduced a general spin-fermion mapping for arbitrary spin
$S$ and spatial dimension that naturally generalizes the Jordan-Wigner
transformation for $S=\frac{1}{2}$. Mathematically, we established a
one-to-one mapping of elements of a Lie algebra onto elements of a
fermionic algebra with a hard-core constraint. Several generalizations,
like a spin-anyon mapping, and important consequences result from these
transformations \cite{cristian}. Incidentally, we note that there are
extremely powerful numerical techniques (cluster algorithms \cite{jim})
to study quantum spin systems, and our mapping allows one to extend
these methods to study the equivalent fermionic problems. 

We thank J.E. Gubernatis for a careful reading of the manuscript.
This work was sponsored by the US DOE under contract
W-7405-ENG-36.

\end{multicols}


\begin{references}
\bibitem{cristian}
More examples and details will be published elsewhere.
\end{references}
\end{document}